# Unified Growth Theory Contradicted by the Economic Growth in Asia

Ron W Nielsen[1]

Environmental Futures Research Institute, Gold Coast Campus, Griffith University, Qld, 4222, Australia

December, 2015

Historical economic growth in Asia (excluding Japan) is analysed. It is shown that Unified Growth Theory is contradicted by the data, which were used (but not analysed) during the formulation of this theory. Unified Growth Theory does not explain the mechanism of economic growth. It explains the mechanism of Malthusian stagnation, which did not exist and it explains the mechanism of the transition from stagnation to growth that did not happen. The data show that the economic growth in Asia was never stagnant but hyperbolic. The alleged dramatic takeoff around 1900 or around any other time did not happen. However, the theory contains also a dangerous and strongly-misleading concept that after a long epoch of stagnation we have now entered the epoch of sustained economic growth, the concept creating the sense of security. The opposite is true. After the epoch of sustained and secure economic growth we have now entered the epoch of a fast-increasing and insecure economic growth.

**Introduction**

In science, data are treated with respect because the primary aim of science is to discover the truth, and for this purpose there is nothing as reliable as a good set data. Many attractive theories and explanations may be formulated but they all have to pass the test of data. Without such a test, they are just stories, which might or might not be true.

It appears that the Unified Growth Theory (Galor, 2005a, 2011) is in this category because it is repeatedly contradicted by data (Nielsen, 2014, 2015a, 2015b, 2015c). In science, one contradicting evidence is sufficient to make the theory, in its original form, unacceptable but the Unified Growth Theory is contradicted more than once. Even more importantly, if not paradoxically, this theory is *contradicted by the same data which were used during its development.* The paradox is easily explained if we notice that Galor did not analyse Maddison's data but was guided solely by impressions created by the customary disfigured representation of data (Ashraf, 2009; Galor, 2005a, 2005b, 2007, 2008a, 2008b, 2008c, 2010, 2011, 2012a, 2012b, 2012c; Galor and Moav, 2002; Snowdon & Galor, 2008).

Galor's predecessors might be excused for believing in the existence of Malthusian stagnation and in the dramatic impact of the Industrial Revolution on changing the economic growth trajectories because these researchers were using strongly limited information. They

---

had no access to the excellent source of data published by the world-renown economist (Maddison, 2001). Galor did.

The latest updated compilation of the economic growth data was published in 2010 (Maddison, 2010) on the year of Maddison's death. This new compilation includes the earlier published data (Maddison, 2001), which were used by Galor, but it extends them to 2008. Any of these compilations can be used to demonstrate that Galor's Unified Growth Theory is repeatedly contradicted by data. The advantage of using the new compilation is to have better information about transitions from the historical hyperbolic growth to slower trajectories, the feature which was ignored in the Unified Growth Theory.

Mathematical analysis (Nielsen, 2015c) of Maddison's data (Maddison, 2010) shows clearly that the historical economic growth, global, regional and national, was hyperbolic and that hyperbolic singularities were repeatedly bypassed by spontaneous diversions to slower, but still-increasing, trajectories. Maddison's data bring a new insight into the interpretation of the historical economic growth. Such a new insight is usually welcome in science. Maddison's data challenge the concepts of Malthusian stagnation, Malthusian trap, escape from Malthusian trap and the dramatic impact of the Industrial Revolution on boosting the economic growth trajectories. It would be unreasonable to suggest that Magnusson deliberately created data to contradict such traditional interpretations.

There appears to be certain reluctance in accepting hyperbolic growth because of its inherent singularity but trajectories can change and Maddison's data demonstrate that hyperbolic growth was systematically changed to slower trajectories at the end of 1980s and the mid-1990s (Nielsen, 2015c). A dramatic example of a change during modern economic growth is Greece where the growth changed from logistic to pseudo-hyperbolic (Nielsen, 2015d). The economic growth in Greece is again changing to a yet-unknown trajectory.

Hyperbolic distributions appear to be causing a problem with their interpretation because they create an impression of being made of two distinctly-different components, slow and fast, joined perhaps by a transition component, the features, which might have convinced Galor and other researchers to accept the doctrine of the existence of two or three different regimes of growth governed by two or three different mechanisms. However, this impression is based on an illusion because hyperbolic distributions are not made of different components. They represent a single, monotonically-increasing growth, which is *impossible* to divide into two or three distinctly-different components.

These issues were discussed elsewhere (Nielsen, 2014). Any person with fundamental knowledge of mathematics will quickly recognise that distributions representing the historical economic growth are hyperbolic. Such a person would quickly realise that the analysis of such distributions is simple because the reciprocal values of hyperbolic distributions are represented by straight lines, and nothing is simpler than the analysis of straight lines.

Hyperbolic growth is described by a simple mathematical formula:

$$S(t) = (a - kt)^{-1} \qquad (1)$$

where, in our case, $S(t)$ is the GDP while $a$ and $k$ are positive constants.

The reciprocal of the hyperbolic distribution is a straight line:

$$\frac{1}{S(t)} = a - kt \qquad (2)$$

Different representations of data are useful in their analysis. It is like using $\ln S(t)$ for exponential distributions. In both cases, more complicated mathematical distributions are converted to linear functions, which are easier to understand. Such conversions can also help in an easy identification of certain types of distributions. For instance, we know that the growth is exponential, or approximately exponential, if $\ln S(t)$ or $\log S(t)$ is linear. Likewise, we know that the growth is hyperbolic if $1/S(t)$ is linear.

Reciprocal values of data can also help in identifying easily any deviations from hyperbolic trend because deviations from a straight line are easy to notice. In using reciprocal values it should be remembered that a deviation to a slower trajectory is indicated by an *upward* bending away from the previous linear trend while deviations to faster trajectories are indicated by *downward* bending. In particular, any form of boosting or a takeoff, repeatedly claimed by Galor for global and regional economic growths, should be indicted by a clear change in the *downward* direction of the reciprocal trajectories.

If the straight line fitting the reciprocal values of data remains undisturbed, it shows that there was no diversion to a faster or slower trajectory. In particular, if the straight line does not show a change in the downward direction (if the gradient of the straight line remains the same) we can conclude that there was no boosting in the economic growth. We obviously cannot claim a change of direction on an undisturbed straight line.

It is impossible to divide a straight line into different sections and claim different mechanism of growth for each of such arbitrarily selected sections. It is impossible to claim, for instance, a transition from stagnation to growth as repeatedly claimed by Galor in his Unified Growth Theory if the reciprocal values of data follow an undisturbed straight line. It is impossible to claim the existence of takeoffs, and it is obviously impossible to claim differential takeoffs if there were no takeoffs. It is also impossible to claim that the Industrial Revolution changed the economic growth trajectory if the reciprocal values of data demonstrate that there was no change, i.e. that their linear trend remained undisturbed.

We shall use the reciprocal values of data in the analysis of the economic growth in Asia but we shall also use the semilogarithmic display of data because in this representation it is easy to study the quality of the hyperbolic fit to the small values of data. Thus, this display can also help to see whether the existence of Malthusian regime could be justified.

**Galor's three regimes of growth**

Before proceeding with the analysis of data for Asia it might be useful to present a brief summary of one of the fundamental postulates of Galor's Unified Growth Theory (Galor, 2005a, 2011), the postulate that the historical economic growth in various countries and regions can be divided into *three distinctly different regimes of growth* governed by distinctly different mechanisms of growth.

These alleged regimes are:

1. The regime of Malthusian stagnation. This regime lasted allegedly for thousands of years and was characterised by random fluctuations and oscillations around a stable Malthusian equilibrium. Galor claims that this epoch of stagnation commenced in 100,000 BC (Galor 2008a, 2012a). Scientific justification for this claim is unclear because Maddison's data (Maddison, 2001), which Galor used (but never analysed) during the formulation of his theory extend only down to AD 1. Extending the alleged epoch of stagnation to 100,000 BC sounds like a large leap of faith. However, the same data, when analysed, demonstrate clearly and convincingly that the three regimes of growth did not exist (Nielsen, 2014) at least for the world economic growth and for the growth in Western Europe. Here we have the theory contradicted not just by one but by two sets of data, which in science is sufficient for the postulate to be rejected. However, Unified Growth Theory is also contradicted by the economic growth in Africa (Nielsen, 2015b).

Galor claims that the regime of Malthusian stagnation was terminated in 1750, or around the time of the Industrial Revolution, 1760-1840 (Floud & McCloskey, 1994), in developed countries and in 1900 in less-developed countries (Galor, 2008a, 2012a). How he managed to determine these dates is unclear because there is already sufficient evidence in Maddison's data (Maddison, 2001, 2010) that Galor's regimes of growth did not exist (Nielsen, 2014, 2015a, 2015b, 2015c).

2. The post-Malthusian regime. According to Galor (Galor, 2008a, 2012a), this mythical regime was between 1750 and 1870 (overlapping the time of the Industrial Revolution) for developed countries but it commenced in 1900, for less-developed countries.

3. The sustained-growth regime. According to Galor (Galor, 2008a, 2012a), this regime commenced in 1870 for developed countries and it still continues.

The claim of different timing for the postulated distinctly different regimes of growth is an integral part of the Unified Growth Theory and is expressed in two other fundamental postulates: the postulate of the differential takeoffs and the postulate of the great divergence, all supported by impressions and by the incorrect interpretation of data. When closely analysed, all these postulated are contradicted by data used by Galor.

The problem with accepting Galor's theory is more than just academic: it has undesirable practical consequences. Galor's concept of the three regimes of growth, the concept he inherited from his many predecessors, is both harmful and misleading. It creates a sense of security when there is none.

According to this concept, after a long stage of economic stagnation, which lasted for many thousands of years, we have now managed to escape from the mythical Malthusian trap and we can, at last, enjoy a sustained economic growth. The opposite is true. The economic growth in the past was not only sustained but also secure (Nielsen, 2015c). However, it is has now reached an insecure stage (Nielsen, 2014, 2015d, 2015e), which requires close monitoring and control. The Unified Growth Theory is not only unscientific (Nielsen, 2014, 2015a, 2015b, 2015c) but also potentially harmful.

**Analysis of data for Asia**

If we examine the list of countries used by Maddison (2010) we can notice that Asia (excluding Japan) is made primarily, if not exclusively, of less-developed countries (BBC, 2014; Pereira, 2011). According to Galor, these countries should have experienced the epoch of stagnation until 1900 followed by the post-Malthusian regime commencing around that year. If Galor's claims are correct, we should see clear signs of stagnation in the data until 1900 and a clear transition (a dramatic takeoff) from stagnation to growth around that year.

Economic growth in Asia between AD 1 and 2008 is presented in Figure 1. There is absolutely no correlation between the data and the three key events indicated in this figure: the Industrial Revolution, the alleged Malthusian regime and the alleged post-Malthusian regime, which were supposed to have been shaping the economic growth.

During the alleged Malthusian regime of stagnation, economic growth in Asia was increasing hyperbolically at least from AD 1000 but the point at AD 1 is also not far away from the calculated hyperbolic distribution. Parameters fitting the data are $a = 2.493 \times 10^{-3}$ and $k = 1.238 \times 10^{-5}$.

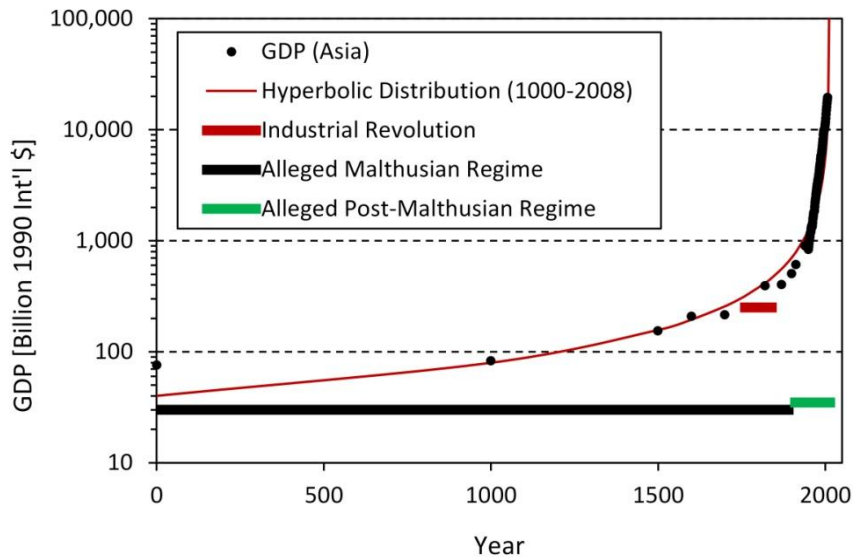

**Figure 1.** Economic growth in Asia (excluding Japan) between AD 1 and 2008. Maddison's data (Maddison, 2010) are compared with the hyperbolic distribution and with their unsubstantiated interpretations promoted by Galor (Galor, 2005a, 2011). The alleged Malthusian regime of stagnation did not exist and neither did the alleged post-Malthusian regime. The growth was hyperbolic from at least AD 1000. The point at AD 1 is 88% higher than the calculated hyperbolic distribution. There was no dramatic transition to a new and faster economic growth after the imaginary epoch of stagnation claimed by Galor, no transition from stagnation to growth claimed repeatedly by Galor (Galor, 2005a, 2011) and no dramatic takeoff. The GDP is in billions of 1990 International Geary-Khamis dollars.

The data show no signs of stagnation, no signs of the Malthusian steady-state equilibrium and no signs of Malthusian oscillations. Assuming the existence of all such features is not only unnecessary but also scientifically unjustified because in science complicated interpretations are rejected in favour of simpler explanations. The data follow a steadily-increasing hyperbolic distribution, suggesting a simple mechanism of growth because hyperbolic distributions are described by a simple mathematical formula [see the eqn (1)].

The data and their analysis give no support to the concept of Malthusian stagnation or to the steady-state Malthusian equilibrium between AD 1 and 1900 or during any other time. It would be incorrect to describe the steadily-increasing growth along the hyperbolic trajectory as stagnation. Such a regular growth suggests the presence of a strong prevailing force. Other random forces might have been present but they must have been averaging out (Kapitza, 2006).

The concept of stagnation is dramatically contradicted by data and so is the transition to the alleged post-Malthusian regime, which was supposed to be a transition from stagnation to growth. We see no such transition but a continuation of the hyperbolic growth. The claimed by Galor takeoff did not happen. There was a minor and hard-to-notice disturbance in the economic growth around 1950 but the growth soon returned to its historical hyperbolic trajectory. The overall evidence in the data is that all these propping-up structures (the alleged different regimes of growth) are redundant and misleading. They can be removed because the data reveal a totally different pattern of growth.

The data and their analysis show that nothing dramatic occurred during the alleged transition from the mythical Malthusian regime of stagnation to the alleged post-Malthusian regime, which is supposed to mark the escape from the mythical Malthusian trap and leading eventually to a sustained growth regime. There was no escape from the trap because there was no trap. During the mythical Malthusian trap the economic growth was steadily increasing and it was obviously unconstrained. It is futile to claim random fluctuations and oscillations when there are none. Why should we even contemplate to make it all more complicated when the data show that the growth was much simpler?

If not for Maddison and his data, the established knowledge in the economic research would have remained established, but now it has to be revaluated and changed. However, new insights should be welcome, particularly if they suggest a simpler explanation of the historical economic growth.

Reciprocal values of the GDP data, 1/GDP, shown in Figure 2, also demonstrate that the Unified Growth Theory is contradicted by the same data, which were used during its development, the data published by Maddison in 2001 (Maddison, 2001) but later extended to include economic growth during the 21st century (Maddison, 2010).

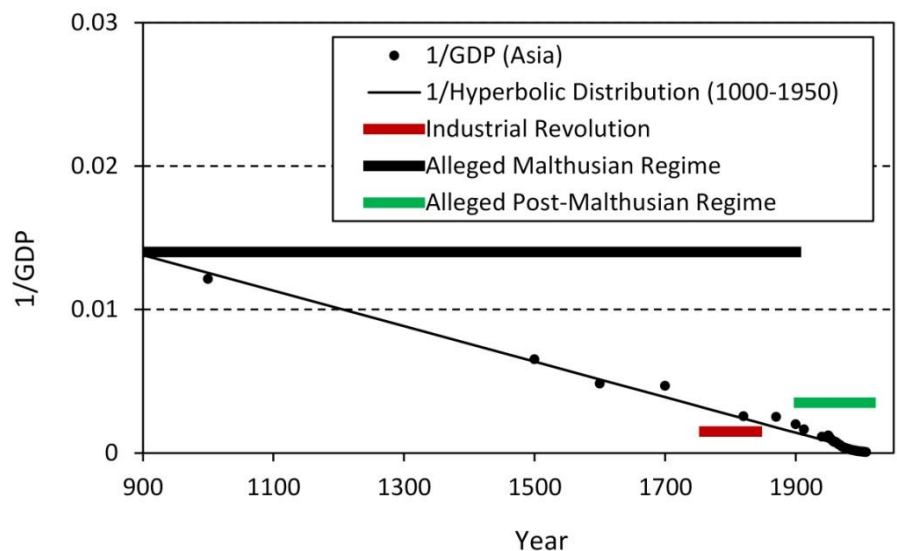

**Figure 2.** Reciprocal values of the GDP data, 1/GDP, for Asia demonstrate that there is no correlation between the claimed events (Industrial Revolution, the alleged Malthusian regime of stagnation and the alleged post-Malthusian regime) and the data (Maddison, 2010). The postulated dramatic and remarkable takeoff around 1900 never happened. The Malthusian regime of stagnation and the post-Malthusian regime did not exist.

During the alleged Malthusian regime of stagnation the reciprocal values of data were decreasing along a straight line indicating an undisturbed, hyperbolic economic growth. The data show also that nothing dramatic had happened at the end of this alleged epoch of stagnation. There was no transition to a new regime of growth. In particular, there was no transition from stagnation to growth, as claimed by Galor, but a continuation of the hyperbolic growth. The concept of the two regimes of growth is convincingly contradicted by data.

**Summary and conclusions**

Analysis of the economic growth in Asia demonstrates that the *Unified Growth Theory (Galor, 2005a, 2011) is contradicted by the same data, which were used (but not analysed) during the formulation of this theory*. Economic growth was increasing along an undisturbed hyperbolic trajectory during and after the alleged regime of Malthusian stagnation. There was no transition from stagnation to growth as claimed by Galor but an undisturbed continuation of the hyperbolic growth. The alleged Malthusian and post-Malthusian regimes did not exist. They represent the redundant and misleading props promoting incorrect interpretations of the mechanism of the economic growth. There was no escape from Malthusian trap because there was no trap. The economic growth in Asia remained steady and unconstrained by any alleged Malthusian limitations. Maddison's data (Maddison, 2001, 2010) also show that there was no transition from stagnation to growth, as claimed by Galor, because the economic growth was never stagnant. A sign of a minor boosting can be noticed around 1950 but it was not a transition from stagnation to growth but a transition from hyperbolic growth to a temporarily slightly faster trajectory, which soon returned to its original hyperbolic trend. Galor's dramatic takeoff never happened.

In science, one contradicting evidence is sufficient to reject contradicted theory or to call for its thorough revision. In the case of the Unified Growth Theory there is more than one evidence. This theory is contradicted by the GDP data describing the world economic growth and the growth in Western Europe (Nielsen, 2014). It is contradicted by the GDP/cap data (Nielsen, 2015a). It is contradicted by the data for Africa (Nielsen, 2015b) and now it is contradicted by the data for Asia. However, implicitly, this theory is also contradicted by the extensive mathematical analysis of the economic growth in various regions and countries (Nielsen, 2015c) showing that the historical economic growth was hyperbolic.

Unified Growth Theory does not explain the mechanism of economic growth. It explains the mechanism of stagnation, which did not exist. It explains the mechanism of transition from stagnation to growth, which never happened. The theory explains features that do not characterise the historical economic growth. It describes phantom features created by impressions and reinforced by the customary crude displays (Ashraf, 2009; Galor, 2005a, 2005b, 2007, 2008a, 2008b, 2008c, 2010, 2011, 2012a, 2012b, 2012c; Galor and Moav, 2002; Snowdon & Galor, 2008) of Maddison's data (Maddison, 2001).

Correct understanding of the historical economic growth is not just an academic issue – it has strong practical implications. Unified Growth Theory, which is strongly based on traditional doctrines, contains a dangerous and misleading concept creating a sense of security when there is none.

According to this theory, after a long epoch of stagnation we have escaped the Malthusian trap, which was imposing strong restrictions on the economic growth, and we have now entered a new epoch of *sustained* economic growth promising the ever-increasing prosperity. At last, after thousands of years of the alleged Malthusian restrictions we have experienced a transition from stagnation to growth. The opposite is true. The past growth was not only sustained but also secure because it was slow and because the generated wealth was well below the ecological limits. The current economic growth is still sustained but unless we take decisive steps to control it, it promises to be unsustainable (Nielsen, 2014, 2015d, 2015e). There is no room for premature celebrations.

If we study global and regional economic growth (Nielsen, 2015c) we shall find there were no transitions from stagnation to growth, as claimed by Galor, but transitions from hyperbolic growth to slower trajectories. However, these slower trajectories continue to increase and they point to an unsustainable economic growth and consequently to an insecure future.

**References**

Ashraf, Q. H. (2009). *Essays on Deep Determinants of Comparative Economic Development*. Ph.D. Thesis, Department of Economics, Brown University, Providence.

BBC, (2014). The North South Divide. http://www.bbc.co.uk/bitesize/standard/geography/international_issues/contrasts_development/revision/2/

Floud, D. & McCloskey, D.N. (1994). *The Economic History of Britain since 1700*. Cambridge: Cambridge University Press.

Galor, O. (2005a). From stagnation to growth: Unified Growth Theory. In P. Aghion & S. Durlauf (Eds.), *Handbook of Economic Growth* (pp. 171-293). Amsterdam: Elsevier.

Galor, O. (2005b). The Demographic Transition and the Emergence of Sustained Economic Growth. *Journal of the European Economic Association*, *3*, 494-504. http://dx.doi.org/10.1162/jeea.2005.3.2-3.494

Galor, O. (2008a). Comparative Economic Development: Insight from Unified Growth Theory. http://www.econ.brown.edu/faculty/Oded_Galor/pdf/Klien%20lecture.pdf

Galor, O. (2008b). Economic Growth in the Very Long Run. In: Durlauf, S.N. and Blume, L.E., Eds., *The New Palgrave Dictionary of Economics*, Palgrave Macmillan, New York. http://dx.doi.org/10.1057/9780230226203.0434

Galor, O. (2008c). Comparative Economic Development: Insight from Unified Growth Theory. http://www.econ.brown.edu/faculty/Oded_Galor/pdf/Klien%20lecture.pdf

Galor, O. (2010). The 2008 Lawrence R. Klein Lecture—Comparative Economic Development: Insights from Unified Growth Theory. *International Economic Review*, *51*, 1-44. http://dx.doi.org/10.1111/j.1468-2354.2009.00569.x

Galor, O. (2011). *Unified Growth Theory*. Princeton, New Jersey: Princeton University Press.

Galor, O. (2012a). Unified Growth Theory and Comparative Economic Development. http://www.biu.ac.il/soc/ec/students/mini_courses/6_12/data/UGT-Luxembourg.pdf

Galor, O. (2012b). The Demographic Transition: Causes and Consequences. *Cliometrica*, **6**, 1-28. http://dx.doi.org/10.1007/s11698-011-0062-7

Galor, O. (2012c). Unified Growth Theory and Comparative Economic Development. http://www.biu.ac.il/soc/ec/students/mini_courses/6_12/data/UGT-Luxembourg.pdf

Galor, O. and Moav, O. (2002). Natural Selection and the Origin of Economic Growth. *The Quarterly Journal of Economics*, *117*, 1133-1191. http://dx.doi.org/10.1162/003355302320935007

Kapitza, S. P. (2006). *Global population blow-up and after*. Hamburg: Global Marshall Plan Initiative.

Maddison, A. (2001). *The World Economy: A Millennial Perspective*. Paris: OECD.

Maddison, A. (2010). Historical Statistics of the World Economy: 1-2008 AD. http://www.ggdc.net/maddison/Historical Statistics/horizontal-file_02-2010.xls.

Nielsen, R. W. (2014). Changing the Paradigm. *Applied Mathematics*, *5*, 1950-1963. http://dx.doi.org/10.4236/am.2014.513188

Nielsen, R. W. (2015a). Unified Growth Theory Contradicted by the GDP/cap Data. http://arxiv.org/ftp/arxiv/papers/1511/1511.09323.pdf

Nielsen, R. W. (2015b). Unified Growth Theory Contradicted by the Economic Growth in Africa, http://arxiv.org/ftp/arxiv/papers/1512/1512.03164.pdf

Nielsen, R. W. (2015c). Mathematical Analysis of the Historical Economic Growth. http://arxiv.org/ftp/arxiv/papers/1509/1509.06612.pdf

Nielsen, R. W. (2015d). Early Warning Signs of the Economic Crisis in Greece: A Warning for Other Countries and Regions. http://arxiv.org/ftp/arxiv/papers/1511/1511.06992.pdf

Nielsen, R. W. (2015e). The Insecure Future of the World Economic Growth. *Journal of Economic and Social Thought*, *2*(4), 242-255.

Pereira, E. (2011). Developing Countries Will Lead Global Growth in 2011, Says World Bank. http://www.forbes.com/sites/evapereira/2011/01/12/developing-countries-will-lead-global-growth-in-2011-says-world-bank/

Snowdon, B. & Galor, O. (2008). Towards a Unified Theory of Economic Growth. *World Economics*, *9*, 97-151.